\documentclass[a4paper]{article}

\usepackage{amsmath,amssymb,epsfig,graphicx,array}
\newcommand{\Ha}{{\mathbf H}}
\newcommand{\M}{{\mathbf M}}
\newcommand{\B}{{\mathbf B}}
\newcommand{\E}{{\mathbf E}}
\newcommand{\D}{{\mathbf D}}
\newcommand{\Po}{{\mathbf P}}
\newcommand{\F}{{\mathbf F}}
\newcommand{\f}{{\mathbf f}}
\newcommand{\pa}{\partial}
\newcommand{\na}{\nabla}

\newcommand{\br}{\boldsymbol{r}}
\newcommand{\ba}{\boldsymbol{a}}
\newcommand{\bn}{\boldsymbol{n}}
\newcommand{\be}{\boldsymbol{e}}
\newcommand{\bj}{\boldsymbol{j}}
\newcommand{\eps}{\varepsilon}

\begin{document}
\noindent
{\Large\bf On the electromagnetic force on a polarizable body}\\[.2cm]

\noindent
A. Engel$^{a)}$ and R. Friedrichs,\\[.1cm]
{\it Institut f\"ur Theoretische Physik, Otto-von-Guericke-Universit\"at
  Magdeburg, PSF 4120, 39016 Magdeburg, Germany}\\[.2cm]

{\small The force on a macroscopic polarizable body in an inhomogenous
  electromagnetic field is calculated for three simple exactly solvable
  situations. Comparing different approaches we pinpoint possible pitfalls
  and resolve recent confusion about the force density in ferrofluids.}

\section{Introduction}

The force on a macroscopic electrically or magnetically
polarizable body in an external electromagnetic field is one of the most
basic and at the same time most relevant problems in classical field
theory. Although all texts $^{1-4}$ 
on the subject introduce Maxwell stress  and the force on a (microscopic)
dipol there is still considerable confusion on how to use these concepts in
concrete situations. That these problems are not confined to the classroom is
exemplified by a recent controversy among experts on the magnetic force
density in ferrofluids $^{5,6}$. 

In order to discuss the possible problems which may arise when calculating the
force on a polarizable body it is very useful to study simple and 
exactly solvable cases. Since a non-zero force requires an {\it inhomogenous}
external field such examples are not abound. The purpose of the present paper
is to present three simple situations in thermodynamic equilibrium for which
fields, stresses and forces can all be calculated exactly, using only basic
Maxwell theory, and to discuss for these situations different methods to
calculate the resulting ponderomotive forces. The first example will be dealt
with in detail whereas for the remaining two only the main results will
be given.

\section{Basic equations}

Neglecting effects of electro- or magnetostriction the Maxwell stress tensor
for a macroscopic polarizable medium in equilibrium is of the form $^{2,4}$ 
\begin{equation}
  \label{eq:defT}
  T_{ij}=-\frac{1}{2}(\eps_0\E^2 +\mu_0\Ha^2)\delta_{ij}+E_i D_j+H_i B_j,
\end{equation}
where as usual $\E$ denotes the electric field, $\D$ the electric
displacement, $\Ha$ the magnetic field, and $\B$ the magnetic induction. From
this stress tensor one finds the Kelvin expression
\begin{equation}
  \label{eq:kelvin}
  f_i:=\pa_j T_{ij}= (P_j \pa_j) E_i + \mu_0 (M_j\pa_j) H_i
\end{equation}
for the ponderomotive force density in the medium. Here $\Po:=\D-\eps_0\E$
denotes the electric polarization and $\M:=\B/\mu_0 -\Ha$ the magnetization of
the medium whereas $\eps_0$ and $\mu_0$ are the dielectric constant and
magnetic permeability respectively of the vacuum. Finally the change of the
equilibrium free energy of a polarizable body due to a change of the fields is
given by $^{3,4}$ 
\begin{equation}\label{eq:defdF}
  \delta F=-\int_V d^3 r\;(\Po\cdot\delta\E_0+ \mu_0\M\cdot\delta\Ha_0),
\end{equation}
where the integral is over the volume of the body, and $\E_0$ and 
$\Ha_0$ denote the external fields in the {\it absence} of the polarizable
material. 

All three expression are consistent with each other and can
therefore be used equally well to determine the total force on a macroscopic
body. In the following examples we will always assume linear constitutive
relations, i.e. $\D=\eps_0 \eps_r \E$ and $\B=\mu_0 \mu_r \Ha$, since we are
only concerned with effects for which a possible nonlinearity in this
relations is not crucial.

\section{Point charge polarizing a half space}

\begin{figure}[htb]\hspace{2.5cm}
  \includegraphics[width=8cm]{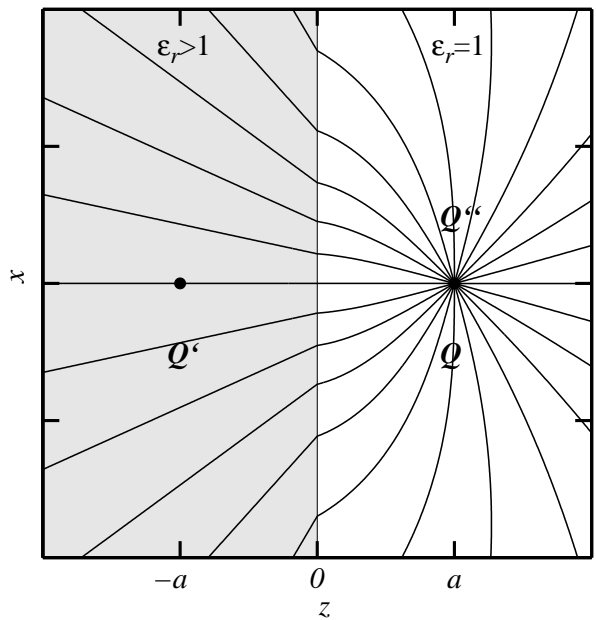}
        \caption{\label{fig1} Lines of the electric displacement field $\D$
          for a point charge $Q$ polarizing a dielectric half space with
          $\eps_r=3$. The field is symmetric with respect to rotations
          around the $z$-axis.}
\end{figure}

Let us first consider a point charge $Q$ at $\ba=(0,0,a)$ a distance $a>0$
away from a dielectric half space with relative dielectric
constant $\eps_r>1$ (see fig.\ref{fig1}). We will neglect possible magnetic
effects and study only the electric part of the problem. 

As is well known $^{2,1}$ the electrostatic problem can be solved
exactly with the help of a mirror charge $Q'=-Q(\eps_r-1)/(\eps_r+1)$ at
$\br=-\ba$ and an additional auxiliary charge $Q''=2Q/(\eps_r+1)$ at
$\br=\ba$. Using the coordinate system of fig.\ref{fig1} the result for the
electric field is 
\begin{equation}
  \label{eq:e1}
  \E(\br)=\left\{\begin{array}{lll}\displaystyle
    \frac{Q}{4\pi\eps_0}\;\frac{2}{\eps_r+1}\;\frac{\br-\ba}{|\br-\ba|^3} &
      \qquad\text{if}\qquad & z\leq 0\\ & & \\ \displaystyle
    \frac{Q}{4\pi\eps_0}\;\left[\frac{\br-\ba}{|\br-\ba|^3}-
      \frac{\eps_r-1}{\eps_r+1}\;\frac{\br+\ba}{|\br+\ba|^3}\right] & 
      \qquad\text{if}\qquad & z\geq 0 
\end{array}\right.\quad.
\end{equation}
Our aim is to determine the force $\F$ with which the point charge attracts 
the half space.

\subsection{The physical picture}
The field of the point charge polarizes the dielectric which results in 
an induced bulk charge density $\rho=-\na \Po$ and a surface charge density 
$\sigma=P_n:=\Po\cdot\bn$ where $\Po=\eps_0(\eps_r-1) \E$ is the
polarization of the medium and $\bn$ denotes the normal vector on the surface
pointing outward of the medium. In our case $\bn=(0,0,1)$. Since the field in
the dielectric is a Coulomb field of a point charge (cf.(\ref{eq:e1})) we
have $\na\Po=0$ and no bulk charge density is induced. For the surface charge
we find from (\ref{eq:e1}) setting $z=0$
\begin{equation}
  \sigma(x,y)=-\eps_0(\eps_r-1)\frac{Q}{4\pi\eps_0}\;
      \frac{2}{\eps_r+1}\;\frac{a}{(x^2+y^2+a^2)^{3/2}}.
\end{equation}
Summing up the Coulomb forces between $Q$ and the elements of this surface
charge we find $\F=(0,0,F)$ with 
\begin{equation}
  \label{eq:F1phys}
  F=\frac{Q^2}{4\pi\eps_0}\;\frac{\eps_r-1}{\eps_r+1}\;\frac{1}{4a^2}.
\end{equation}
This is an unambigous and physically sound expression for the force in
question. It is exactly equal to the Coulomb force between $Q$ and
the mirror charge $Q'$. 

\subsection{Integrating the Kelvin force}
For the Kelvin force density (\ref{eq:kelvin}) we find in the present
situation  
\begin{equation}
  \label{eq:kelvin1}
  \f=\frac{\eps_0}{2}(\eps_r-1)\na \E^2=
     -\frac{Q^2}{(4\pi\eps_0)^2}\;\frac{8\eps_0(\eps_r-1)}{(\eps_r+1)^2}\;
           \frac{\br-\ba}{|\br-\ba|^6}.
\end{equation}
Integrating this expression over the halfspace $z<0$ the $x$ and $y$
components again average to zero and for the $z$ component we get 
\begin{equation}\label{eq:F1kelvin}
  F^{\text{Kelvin}}=\frac{Q^2}{4\pi\eps_0}\;
     \frac{2(\eps_r-1)}{(\eps_r+1)^2}\;\frac{1}{4a^2}.
\end{equation}
This is clearly different from (\ref{eq:F1phys}): the integral over the Kelvin
force density does {\it not} yield the total force on a polarizable body. This
is what was observed also in $^5$ for another example.
The reason for this discrepancy is however not that the
expressions (\ref{eq:kelvin}) or (\ref{eq:kelvin1}) for the Kelvin force
density are incorrect but that the {\it total} electromagnetic force on the
polarizable medium involves both bulk and surface contributions whereas
(\ref{eq:F1kelvin}) only covers the bulk part. The expression for the
appropriate surface part is given below.

\subsection{Balance of stresses}

By the definition of stress the total force on a body is given by the integral
of the stress tensor over the complete surface of the body. Paying attention
to the fact that the Maxwell stress is non-zero also outside the polarizable
medium a calculation of the electromagnetic force on a body has to integrate
the {\it differences} in stress slightly above and below the surface of the
body. In our particular problem the fields
decay sufficiently rapidly to zero for $r\to\infty$ such that only the
integral over the $x$-$y$-plane contributes. Using (\ref{eq:defT}) and
(\ref{eq:e1}) we find for the relevant component of the Maxwell stress tensor 
\begin{equation}\label{eq:Tel1}
  \lim_{z\to-0}T_{zz}(x,y,z)=\frac{Q^2}{(4\pi\eps_0)^2}\;
    \frac{2\eps_0}{(\eps_r+1)^2}\left[
          \frac{2\eps_r\; a^2}{(x^2+y^2+a^2)^3}
          -\frac{1}{(x^2+y^2+a^2)^2}\right]
\end{equation}
on the medium side and 
\begin{equation}\label{eq:Telvac1}
  \lim_{z\to+0}T_{zz}(x,y,z)=\frac{Q^2}{(4\pi\eps_0)^2}\;
    \frac{2\eps_0}{(\eps_r+1)^2}\left[
    \frac{\eps_r^2\; a^2}{(x^2+y^2+a^2)^3}
    -\frac{x^2+y^2}{(x^2+y^2+a^2)^3}\right]
\end{equation}
on the vacuum side. Integrating the difference between these electric stresses
one finds for the $z$ component of the force
\begin{equation}
  \label{eq:fel1}
  F^{\text{electric}}=\frac{Q^2}{4\pi\eps_0}\;
     (\frac{\eps_r-1}{\eps_r+1})^2\;\frac{1}{4a^2}.
\end{equation}
This is again different from the correct expression (\ref{eq:F1phys})! The
reason for this renewed failure is somewhat more subtle. The point is that we
have only considered the {\it electric} part of the stress tensor in the
medium whereas the {\it total} force must clearly be related to the surface
integral of the {\it total} stress tensor $T^{\text{total}}_{ij}$. But in
equilibrium we must have $\pa_j T^{\text{total}}_{ij}=0$ (otherwise there
would be momentum transport in the medium) and the electric part
(\ref{eq:Tel1}) is hence counterbalanced 
by non-electric contributions. If our dielectric is a solid, elastic stresses
would build up, if it is a fluid a non-trivial pressure field would compensate
the electric stress. In any case the total internal stress in equilibrium must
be divergence free and hence cannot contribute to the total force on the body.

The whole stress on the surface is therefore given by the vacuum part
(\ref{eq:Telvac1}) and its surface integral indeed yields the correct result 
(\ref{eq:F1phys}). Incidentally this also explains why the force is equal to
the Coulomb force between $Q$ and $Q'$: the field for $z>0$ is identical for
either the polarizable half space or the dielectric replaced by the mirror
charge $Q'$ and hence the integrals over the vacuum stress must be the same.

Let us finally note that from the definition $f_i:=\pa_j T_{ij}$ of the Kelvin
force density it is clear that the volume integral of $f_i$ must be identical
to the surface integral of $T_{ij}$. Hence the elimination of the contribution
from the electric stress inside the medium in (\ref{eq:fel1}) can simply be 
accomplished by adding (\ref{eq:F1kelvin}). Equivalently (\ref{eq:fel1}) is
nothing but the surface contribution that was missed in
(\ref{eq:F1kelvin}). Using the behaviour of the fields $\E$ and $\D$ at the
interface between media with different dielectric constant it is easy to show
$^4$ that 
\begin{equation}
  n_i(T_{ij}^{\text{vacuum}}-T^{\text{medium}}_{ij})n_j=
                    \frac{1}{2\eps_0} (\Po\cdot\bn)^2.
\end{equation}
Hence the appropriate surface force density complementing the Kelvin body
force density is given by 
\begin{equation}
  \f^{\text{surface}}=\frac{1}{2\eps_0} \Po_n^2 \;\bn.
\end{equation}

\subsection{Change in free energy}

The simplest and most direct way to determine the total force on a
polarizable body in equilibrium is to use expression (\ref{eq:defdF}) for the
variation in the free energy. Specializing to the case without magnetic
contributions we note first that the change in the electric field $\E_0$ in
the region of the body 
due to a displacement of the body by an infinitesimal vector $\delta \br$ is 
given by $\delta\E_0=(\delta \br\cdot\na)\;\E_0$. On the other hand the
corresponding change in free energy is related to the force $\F$ via $\delta F
= -\F\cdot\delta \br$ by the very definition of the total force $\F$. Using
the condition $\na\times\E_0=0$ fulfilled in thermodynamic
equilibrium we therefore find  
\begin{equation}\label{eq:main}
  \F=\int_V d^3 r \;(\Po\cdot\na)\;\E_0.
\end{equation}
Using (\ref{eq:e1}) to determine $\Po$ as well as  
\begin{equation}
  \E_0(\br)=\frac{Q}{4\pi\eps_0}\;\frac{\br-\ba}{|\br-\ba|^3}
\end{equation}
and performing the integral we indeed recover the correct expression 
(\ref{eq:F1phys}).

\section{Charged line polarizing a parallel cylinder}

\begin{figure}[htb]\hspace{2.5cm}
  \includegraphics[width=8cm]{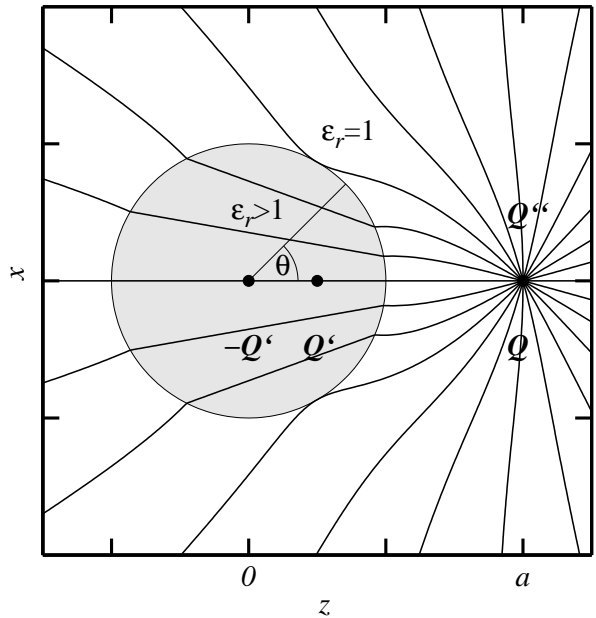}
        \caption{\label{fig2} Lines of the electric displacement field $\D$
          for a charged line polarizing a parallel dielectric cylinder with
          $\eps_r=3$ and radius $R=a/2$, where $a$ denotes the distance
          between the axis of the cylinder and the charged line. The situation
          is symmetric with respect to translations in the $y$-direction.}
\end{figure}

This is again an example from electrostatics. A line at distance $a$ from and
parallel to the $y$-axis carries a charge $Q$ per length. A dielectric
cylinder of radius $R$, dielectric constant $\eps_r$, and the $y$-axis as
symmetry axis is polarized by the field of the line charge (see
fig.\ref{fig2}). We want to determine the force with which the cylinder is
attracted to the line. 

Due to the translational invariance along the $y$-axis the problem is
essentially two-dimensional and accordingly we use the notation $\br=(x,0,z)$
in the following. The electrostatic problem can again be solved  
by the method of images. To determine the fields outside
the dielectric one introduces two additional charged lines with densities
$Q'=-Q(\eps_r-1)/(\eps_r+1)$ at a distance $a':=R^2/a$ from the $y$-axis and
$-Q'$ at the $y$-axis respectively. The field inside the dielectric is
determined by replacing the original line by one with charge density 
$Q''=2Q/(\eps_r+1)$. The resulting electric field is given by 
\begin{equation}\label{eq:e2}
  \E(\br)=\left\{\begin{array}{lll}\displaystyle
    \frac{Q}{2\pi\eps_0}\;\frac{2}{\eps_r+1}\;\frac{\br-\ba}{|\br-\ba|^2} &
      \text{if} & r\leq R\\ & & \\ \displaystyle
    \frac{Q}{2\pi\eps_0}\;\left[\frac{\br-\ba}{|\br-\ba|^2}-
      \frac{\eps_r-1}{\eps_r+1}\;\frac{\br-\ba'}{|\br-\ba'|^2}
    +\frac{\eps_r-1}{\eps_r+1}\;\frac{\br}{r^2}\right] & 
      \text{if} & r\geq R
\end{array}\right.\quad.
\end{equation}
The field inside the dielectric is again divergence free. The induced surface
charge can be determined from the normal component of the polarization as in
the first example and is found to be 
\begin{equation}\label{eq:sigma2}
  \sigma(\theta)=\frac{Q}{\pi}\;\frac{\eps_r-1}{\eps_r+1}\;
       \frac{R-a\cos\theta}{R^2-2Ra\cos\theta +a^2},
\end{equation}
where $\theta$ denotes the angle between the radius vector to the surface
element in question and the $z$-axis. Using the two-dimensional Coulomb law
and summing up the $z$-components of the forces between the surface elements
and the original charged line we find 
\begin{equation}\label{eq:force2}
 F=\frac{Q^2}{2\pi\eps_0}\;\frac{\eps_r-1}{\eps_r+1}\;
        \frac{R^2}{a(a^2-R^2)}.
\end{equation}
As expected this is exactly the sum of the Coulomb forces between $Q, Q'$ and
$-Q'$. 

The most direct way to calculate this force is to use (\ref{eq:defdF}) in the
form (\ref{eq:main}). Since $\Po$ is parallel to $\E_0$ we have 
\begin{equation}
 (\Po \na)\E_0=P\na E_0=-\frac{Q^2}{2\pi^2\eps_0}\;\frac{\eps_r-1}{\eps_r+1}\;
          \frac{\br-\ba}{|\br-\ba|^4}.
\end{equation}
Integrating this expression over the volume of the dielectric cylinder the
$x$-component averages to zero whereas the $z$-component yields back
(\ref{eq:force2}). 

\section{Magnetizable cylinder in the field of a straight wire}

Finally we discuss a simple magnetic example for which the electric part will
be neglected. Consider a straight wire at distance $a$ from the $y$-axis
carrying a current $I$. A cylinder of radius $R$ and magnetic permeability
$\mu_r$ with the $y$-axis as symmetry axis is
placed into the magnetic field of the wire. We want to determine the force per
length on the cylinder. 
\begin{figure}[htb]\hspace{2.5cm}
  \includegraphics[width=8cm]{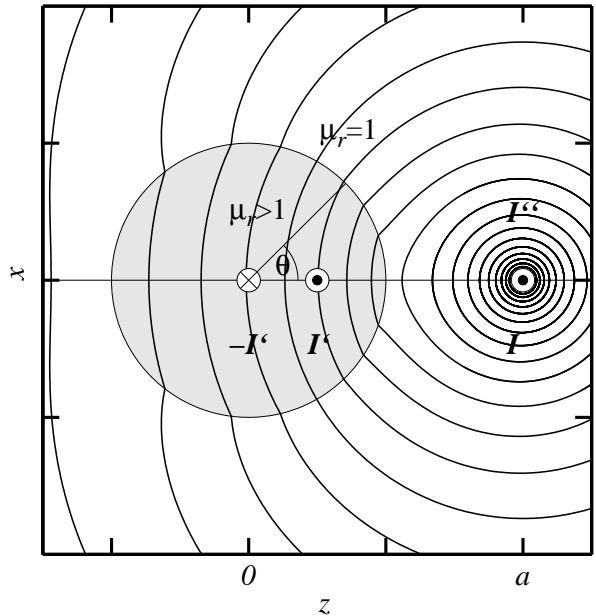}
        \caption{\label{fig3} Lines of the magnetic induction $\B$
          for a current-carrying straight wire magnetizing a parallel cylinder
          with magnetic permeability $\mu_r=3$. The geometry is identical with
          fig.\ref{fig2}.} 
\end{figure}

Due to the well known similarities between the magnetic fields of parallel,
stationary currents and two-dimensional electrostatics the algebra for this
example is very similar to the previous one. We introduce two mirror currents,
$I'=I(\mu_r-1)/(\mu_r+1)$ parallel to the original one at a distance
$a'=R^2/a$ from the $y$-axis and $-I'$ along the
$y$-axis. The field inside the medium is given by replacing $I$ by
$I''=2I/(\mu_r+1)$. The magnetic field is then found to be 
\begin{equation}\label{eq:m1}
  \Ha(\br)=\left\{\begin{array}{lll}\displaystyle
 \frac{I}{2\pi}\;\frac{2}{\mu_r+1}\;
      \frac{\be_y\times(\br-\ba)}{|\br-\ba|^2} &
      \text{if}& r\leq R\\ & & \\ \displaystyle
\frac{I}{2\pi}\;\left[\frac{\be_y\times(\br-\ba)}{|\br-\ba|^2}+
      \frac{\mu_r-1}{\mu_r+1}\;\frac{\be_y\times(\br-\ba')}{|\br-\ba'|^2}
    -\frac{\mu_r-1}{\mu_r+1}\;\frac{\be_y\times\br}{|\br|^2}\right] & 
      \text{if}& r\geq R
\end{array}\right.,
\end{equation}
where $\be_y$ denotes the unit vector in $y$-direction. 

To clarify the physical picture we note that 
the magnetization $\M=(\mu_r-1)\Ha$ correspondes to induced bulk currents of
density $\bj_b=\na\times \M$ and induced surface currents
$\bj_s=\M\times\bn$. In our case $\na\times \M=0$ and 
\begin{equation}
  \bj_s(\theta)=\frac{I}{\pi}\;\frac{\mu_r-1}{\mu_r+1}\;
       \frac{a\cos\theta-R}{R^2-2Ra\cos\theta +a^2}\be_y ,
\end{equation}
which is analogous to (\ref{eq:sigma2}). Correspondingly summing up the
magnetic forces between these surface currents and the original one we find 
\begin{equation}\label{eq:force3}
 F=\frac{\mu_0 I^2}{2\pi}\;\frac{\mu_r-1}{\mu_r+1}\;
        \frac{R^2}{a(a^2-R^2)}.
\end{equation}
This result is most easily derived  by using (\ref{eq:defdF}) which in the
present case gives rise to 
\begin{equation}
  \F=\mu_0\int_V d^3 r \;(\M\cdot\na)\;\Ha_0.
\end{equation}
Exploiting the fact that $\M$ and $\Ha_0$ are parallel it is easy to reproduce
(\ref{eq:force3}).

\section{Summary}

Three simple examples have been presented for which the
equilibrium force on a polarizable body in an external electromagnetic field
can be determined exactly. Different ways to do this have been elucidated. A
first possibility is to integrate the Maxwell stress tensor (\ref{eq:defT})
over a surface {\it outside} the body and completely enclosing it. The second
method integrates the volume force density (\ref{eq:kelvin}) and 
{\it adds the surface integral} over the difference in normal stresses 
$(P_n^2/(2\eps_0)+\mu_0 M_n^2/2)$ involving the normal components of the
polarization and the magnetization. The third and in our opinion most simple
and straightforward method builds on the free energy change (\ref{eq:defdF})
and uses the formula
\begin{equation}
    \F=\int_V d^3 r \;\left[(\Po\cdot\na)\;\E_0
            +\mu_0 (\M\cdot\na)\;\Ha_0\right],
\end{equation}
where $\E_0$ and $\Ha_0$ denote the fields in the absence of the polarizable
medium. Note that in the case where the fields do not vary much over the volume
of the body this expression reduces to the well known relation
\begin{equation}
    \F=({\boldsymbol{\cal P}}\cdot\na)\;\E_0+
       \mu_0 ({\boldsymbol{\cal M}}\cdot\na)\;\Ha_0,
\end{equation}
with ${\boldsymbol{\cal P}}$ and ${\boldsymbol{\cal M}}$ denoting the 
{\it total} polarization and magnetization respectively of the body.\\[.5cm]

\noindent
{\bf Acknowledgement:} We would like to thank Adrian Lange and Hanns-Walter
M\"uller for a critical reading of the manuscript. \\[.5cm]

\noindent
$^{a)}$ andreas.engel@physik.uni-magdeburg.de\\
$^1$ John D. Jackson, {\it Classical Electrodynamics} (John Wiley
  \& Sons, New York, 1975)\\
$^2$ Richard Becker and Fritz Sauter, {\it Theorie der
    Elektrizit\"at} (Teubner, Stuttgart, 1973)\\
$^3$ Julius A. Stratton, {\it Electromagnetic Theory}
  (McGraw-Hill, New York, 1941)\\
$^4$ L. D. Landau, E. M. Lifshitz, {\it Lehrbuch der theoretischen
    Physik, Band VIII, Elektrodynamik der Kontinua} (Akademie-Verlag, Berlin,
  1980)\\
$^5$ Stefan Odenbach and  Mario Liu, Phys. Rev. Lett. {\bf 86}, 328
  (2001) \\
$^6$ comments on $^5$, to appear in Phys. Rev. Lett.

\end{document}